\documentclass{cimento}

\usepackage{graphicx}  

\title{Recent insights on the penumbra formation process}
\author{M.~Murabito\from{ins:a}\from{ins:b}\ETC,
P.~Romano\from{ins:c},
F.~Zuccarello\from{ins:a},
S.L.~Guglielmino\from{ins:a}}
\instlist{\inst{ins:a} Dipartimento di Fisica e Astronomia -- Sez.~Astrofisica, Universit\`{a} degli Studi di Catania, Via S.~Sofia 78, I-95123 Catania, Italy
	\inst{ins:b} INAF -- Oss.~Astronomico di Roma, via Frascati 33, I-00078 Monte Porzio Catone, Italy
	\inst{ins:c} INAF -- Oss.~Astrofisico di Catania, Via S.~Sofia 78, I-95123 Catania, Italy}

\begin{document}

\maketitle

\begin{abstract}

Using high-resolution spectropolarimetric data acquired by \textit{IBIS}, as well as \textit{SDO}/HMI observations, we studied the penumbra formation in AR NOAA 11490 and in a sample of twelve ARs appeared on the solar disk on 2011 and 2012, which were characterized by $\beta$-type magnetic field configuration. 
The results show that the onset of the classical Evershed flow occurs in a very short time scale, 1-3 hours. Studying the formation of the first penumbral sector around the following proto-spot, we found that a stable penumbra forms in the area facing the opposite polarity, which appears to be co-spatial with an AFS, i.e. in a flux emergence region, in contrast with the results of \cite{Schlichenmaier2010} concerning the leading polarity of AR NOAA 11490. 
Conversely, analyzing the sample of twelve ARs, we noticed that there is not a preferred location for the formation of the first penumbral sector. We also observed before the penumbra formation an inverse Evershed flow, which changes its sign when the penumbra appears. This confirms the observational evidence that the appearance of the penumbral filaments is correlated with the transition from the inverse Evershed to the classical Evershed flow. Furthermore, the analysis suggests that the time needed to form the penumbra may be related to the location where the penumbra first appears. New high-resolution observations, like those that will be provided by the European Solar Telescope, are expected to increase our understanding of the penumbra formation process.
 
\end{abstract}

\section{Introduction}

Penumbra formation is a physical process which involves the coupling between plasma and magnetic field in the solar atmosphere. Some aspects concerning this process remain unclear due to few available datasets of this phase of the sunspot evolution.

Recently, \cite{Schlichenmaier2010} found that the penumbra grows sector by sector, developing on the side facing the opposite polarity of the forming active region (AR). The area between the two polarities characterized by elongated granules associated with flux emergence. This confirmed numerical simulations (\cite{Cheung2008,Tortosa2009}), which also suggested that such dynamics prevents the settlement of a stable penumbra. Moreover, \cite{Schlichenmaier2010} observed for the first time a line-of-sight (LOS) velocity of opposite sign with respect to that displayed by the typical Evershed flow (EF, \cite{Evershed1909}) at some azimuths around the sunspot, before the penumbra formation.

In this contribution we report a summary of our analyses concerning the penumbra formation (\cite{Murabito:16,Murabito:17,Murabito:18}), focusing on both the onset of the Evershed flow and on the location of the primary site where the first stable penumbral sector appears.

\section{Observations and Data Analysis}

AR NOAA 11490 was observed by the spectropolarimeter \textit{IBIS} (\cite{Cav06}) mounted on the Dunn Solar Telescope, in New Mexico, on 2012 May 28 and 29. 

\begin{figure}[b]
	\centering
	\includegraphics[trim=0 180 200 285, clip, scale=0.25]{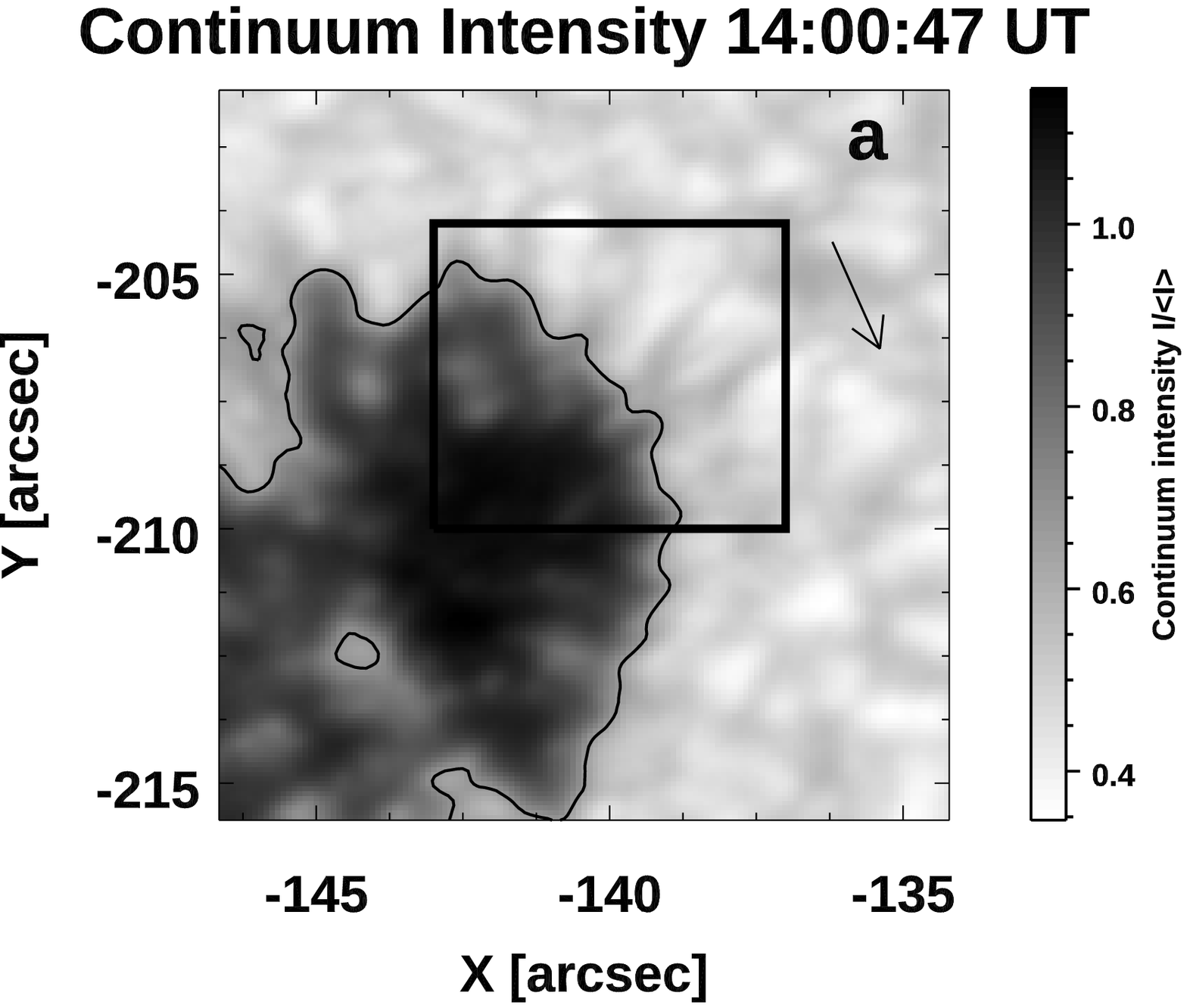}
	\includegraphics[trim=30 180 100 285, clip, scale=0.25]{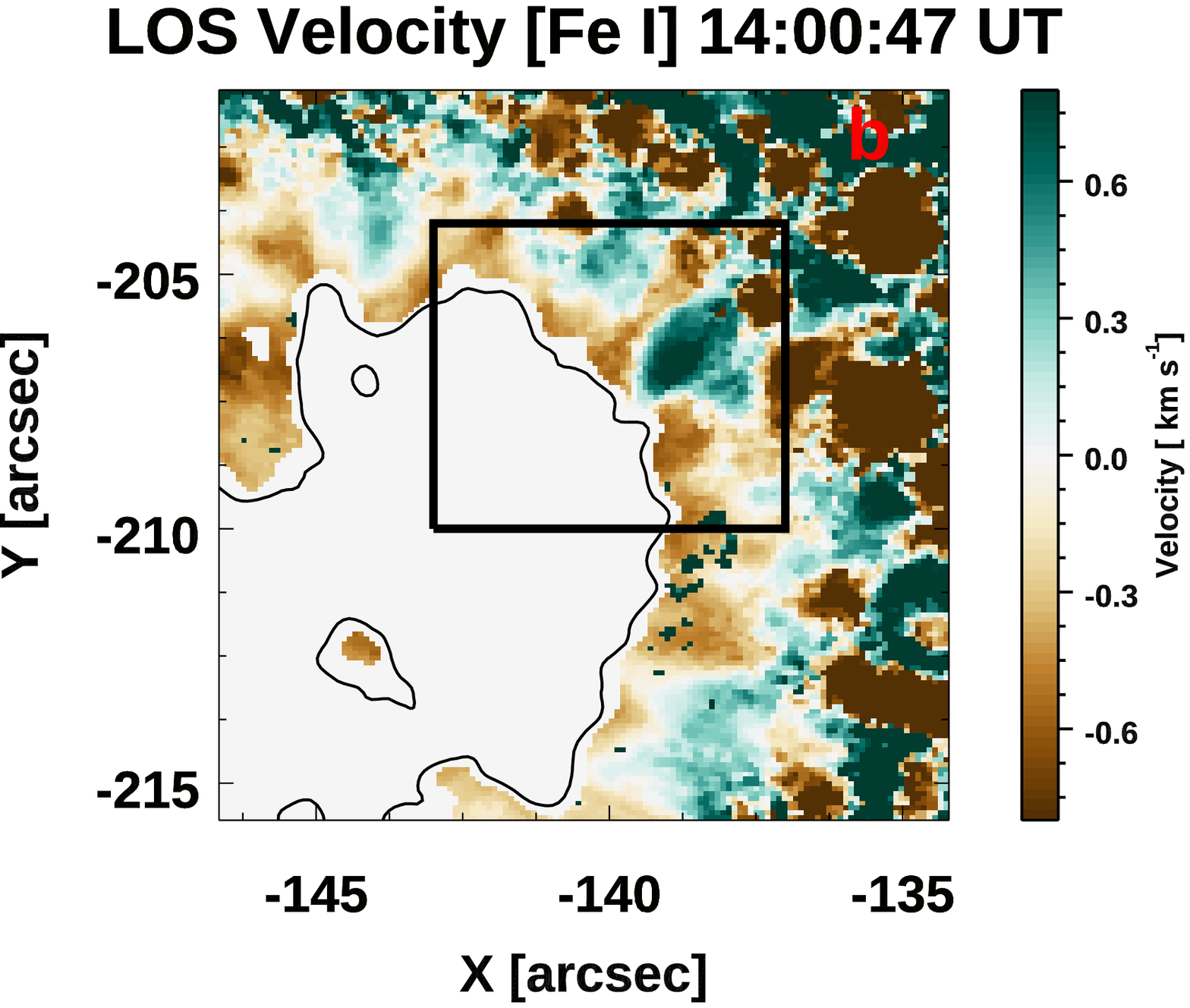}
	\includegraphics[trim=50 0 20 100, clip,scale=0.2]{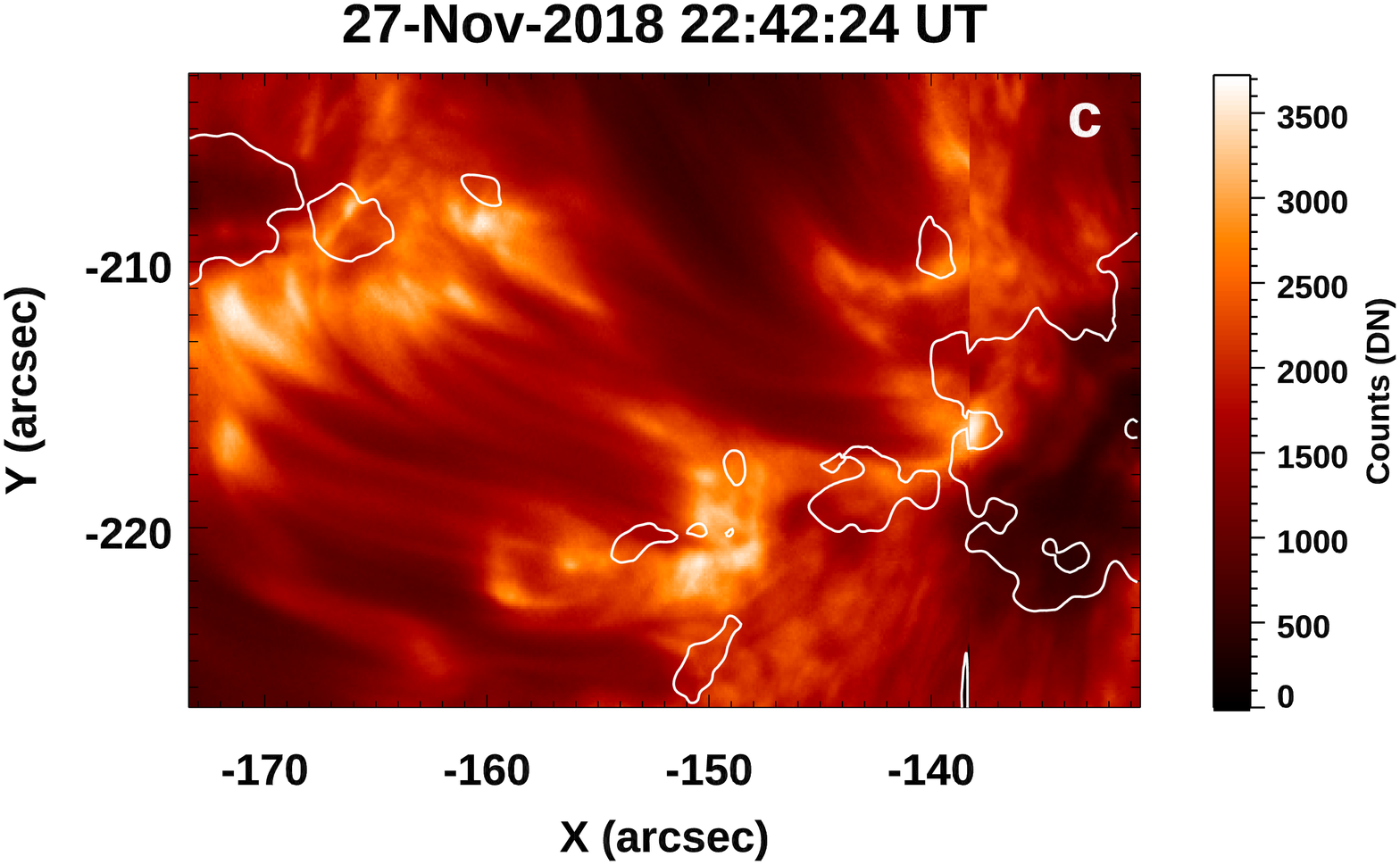}
	\caption{Zoom of intensity and LOS velocity (\textit{a - b}) maps of the preceding spot of AR NOAA 11490. Intensity map in the center of the Ca II 854.2 nm line of the whole AR(\textit{c}) . \label{fig:1} }
\end{figure}

Firstly, we analyzed the proceding spot of the AR by using both photospheric (617.3~nm line) \textit{IBIS} and \textit{SDO}/HMI measurements (\cite{Murabito:16}). The pore, before the penumbra formation (Figure~\ref{fig:1}a), exhibits an annular zone where the magnetic field is around 1~kG and has an \textit{upside down ballerina skirt} structure, as already showed in the analysis of \cite{Rom13, Rom14}. As shown in Figure~\ref{fig:1}(b), the annular zone is characterized by downflows. This analysis was focused on the transition from this inverse EF to the classical one, which occurs in $1-3$ hours.

In the subsequent analysis (\cite{Murabito:17}), we studied the penumbra formation of the following spot of the AR, showing that the first penumbral sector appears on the side facing the opposite polarity. We used the photospheric and chromospheric (854.2~nm line) \textit{IBIS} data as well as the \textit{SDO}/HMI magnetograms. Elongated granules in photosphere, an AFS in chromosphere (see Figure~\ref{fig:1}(c)) and a sea-serpent configuration of the magnetic field are visible in the penumbra formation area.

\begin{figure}[t]
	\centering
	\includegraphics[trim=180 300 230 410, clip, scale=0.57]{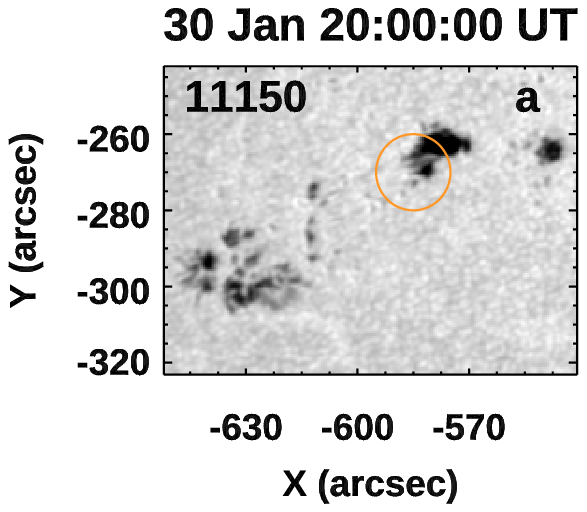}
	\includegraphics[trim=210 300 230 410, clip, scale=0.57]{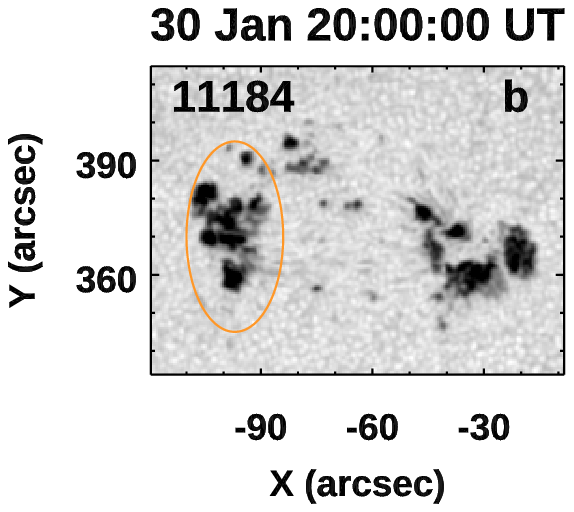}
	\includegraphics[trim=210 300 230 410, clip, scale=0.57]{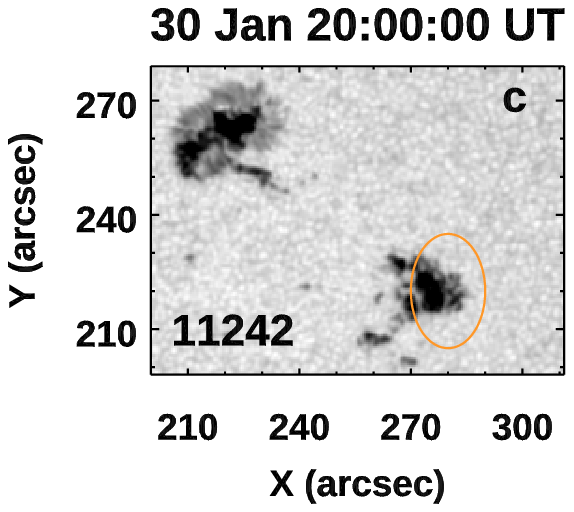}
	\includegraphics[trim=210 300 230 415, clip, scale=0.57]{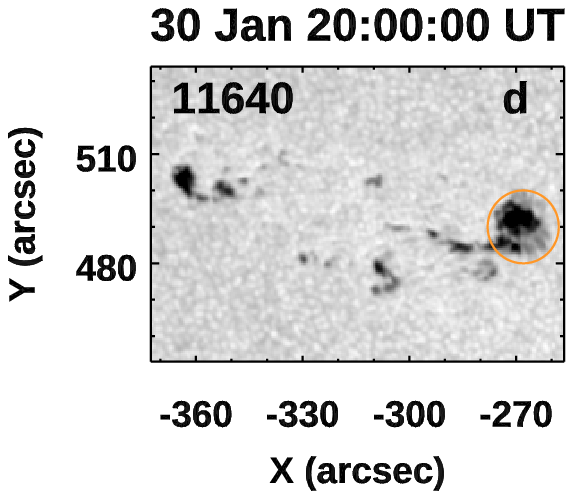}
	\caption{\textit{SDO}/HMI continuum filtergrams showing samples of the twelve selected ARs. Panels \textit{a and b} show spots classified as B-type spots and panels \textit{c and d} spots labelled A-type spots. (All the images are adapted from \cite{Murabito:16}, \cite{Murabito:17} and \cite{Murabito:18}) \label{fig:2} }
\end{figure}

Finally, considering twelve ARs, observed on 2011 and 2012 during the maximum of solar cycle 24 with \textit{SDO}/HMI, and classifying them according to whether the first penumbral filaments appear on the side away (A-type) or facing the opposite polarity (B-type), we highlighted that there is no a preferred location for the penumbra formation (\cite{Murabito:18}). We shown four different examples in Figure~\ref{fig:2}. The penumbra in the two spots displayed in Figure~\ref{fig:2}(a-b) forms between the two polarities, while in the other two spots in Figure~\ref{fig:2}(c-d) the first penumbral sector appears in the region away the opposite polarity.

\section{Conclusions}
High resolution observations of AR NOAA 11490 indicate two different behaviours for the penumbra formation process. The preceding spot forms the penumbra sector by sector, in agreement with \cite{Schlichenmaier2010}, the last sector appearing on the side facing the opposite polarity. The first stable penumbral sector around the
following spot forms along the side between the two polarities, where elongated granules, filamentary magnetic field in the photosphere and an AFS in the chromosphere are visible, in contrast to the results of \cite{Schlichenmaier2010}. 
The analysis of a sample of ARs shows that there is no a preferred location for the formation of the first penumbral sector. The time needed for the penumbra formation may be related to the location where the first penumbral sector appears \cite{Murabito:18}.

Concerning the onset of the EF, we found that it occurs in a short time scale, 1-3 hrs. The sample of ARs observed by \textit{SDO}/HMI provides evidence that there is a correlation between the appearance of the penumbral filaments and the transition from the inverse to the classical EF. Independently on the type and stage of the penumbra formation, the flow changes its sign when the penumbra appears. This confirms the scenario proposed in \cite{Murabito:16}.

\acknowledgments
This work has received funding from the European Commission's Seventh Framework programme under the grant agreement SOLARNET (project no.~312495). This work was also supported by the Istituto Nazionale di Astrofisica (PRIN-INAF-2014), by the University of Catania (Linea di intervento 2) and by Space Weather Italian COmmunity (SWICO) Research Program.

\end{document}